\documentclass[%
reprint,
 amsmath,amssymb,
 aps,
 pra,
 ]{revtex4-1}
\usepackage[T1]{fontenc}
\usepackage{graphicx}
\usepackage{dcolumn}
\usepackage{bm}
\usepackage{geometry}
\usepackage{color}
\begin{document}
\preprint{APS/123-QED}
\title{Laser ablation loading of a radiofrequency ion trap}
\author{K. Zimmermann, M. V. Okhapkin, O. A. Herrera-Sancho, and E. Peik}

 \email{Corresponding author E-mail address: ekkehard.peik@ptb.de}
 \affiliation{Physikalisch-Technische Bundesanstalt, Bundesallee 100, 38116 Braunschweig, Germany}

\date{\today}

\begin{abstract}
The production of ions via laser ablation for the loading of radiofrequency (RF) ion traps is investigated using a nitrogen laser with a maximum pulse energy of 0.17 mJ and a peak intensity of about 250 MW/cm$^2$. A time-of-flight mass spectrometer is used to  measure the ion yield and the distribution of the charge states. Singly charged ions of elements that are presently
considered for the use in optical clocks or quantum logic applications could be produced from metallic samples at a rate of the order of magnitude  $10^5$ ions per pulse.
A linear Paul trap was loaded with Th$^+$ ions
produced by laser ablation. An overall ion production and trapping efficiency of $10^{-7}$ to $10^{-6}$ was attained.
For ions injected individually, a dependence of the capture probability on the phase of the RF field has been predicted. In the experiment this was not observed, presumably because of collective effects within the ablation plume.
\end{abstract}

\maketitle

\section{Introduction}\label{sec:intro}

Experiments with ions in radiofrequency Paul traps or Penning traps are usually prepared by producing  ions with low kinetic energy inside the trapping volume. For
non-volatile species, originally a combination of a small atomic beam oven with an electron gun was used to ionize atoms by electron impact
\cite{fischer,neuhauser}. This technique has two disadvantages: the atomic beam may unintentionally deposit an inhomogeneous layer of the
element to be trapped on the trap electrodes, thus creating an inhomogeneous surface potential. The electron beam may leave stray charges on the surface of
insulators situated in the vicinity of the trap. Both effects make the precise and stable localization of the ion in the trap more tedious. These
adverse effects can be strongly mitigated by using resonant laser photoionization of the neutral atom \cite{drewsen1,blatt}, allowing to eliminate
the electron beam and to reduce the atomic flux because of the higher ionization cross section. This approach adds to the complexity of the setup,
however, because one or even two lasers are required to excite resonance transitions of the neutral atom.

Here we present a systematic study of ion production via photoablation using a nitrogen laser.
This type of laser was selected because it provides the most compact source of nanosecond pulses with a peak power of tens of kW  and because its wavelength of 337~nm will generally be absorbed more strongly in metals than the infrared radiation available from other efficient pulsed lasers.
The motivation for this study was to establish a trap loading method that requires less source material and produces less stray charges than electron impact ionization. The advantage of laser ablation results from the fact that a small source volume is heated rapidly to obtain the required vapor pressure over a very short time interval. In comparison with resonant photoionization, laser ablation loading is technically less complex. Ideally, laser ablation and photoionization may be combined, as has been investigated for Ca$^+$ \cite{hendricks,sheridan}, because this provides isotope-selective ion production with a minimum amount of stray charge. In our experiments, ions of most elements that are presently
considered for the use in optical clocks or quantum logic applications were produced effectively from metallic samples. This method
also allows one to produce ions of elements that are not easily evaporated like e.g. tungsten or thorium and it also produces doubly charged ions.
We are especially interested in the production of Th$^+$ for an investigation of a nuclear optical clock based on the low energy transition to an isomeric state in $^{229}$Th \cite{porsev}.  Previous
studies have already reported laser ablation from metals for the loading of ion traps \cite{knight,kwong1,hashimoto,leib,kwapien,hendricks,chapman}, but typically higher laser pulse energies at infrared
wavelengths were used and experiments were carried out only for a small selection of elements. Laser desorption of organic
molecules has been studied in conjunction with ion trap mass spectrometry (see for example Refs. \cite{hemberger,robb}).

Photoablation is categorized by two different regimes which are distinguished by comparing the laser pulse duration with the
characteristic time of electron-phonon interaction. When the laser pulse duration is shorter than this characteristic time, which is typically several picoseconds in metals, bonds may be broken by the strong electric field and ablation can be considered as a direct solid-to-vapor transition.  In contrast, in the interaction of
metals with nanosecond laser pulses as investigated in this paper, the pulse duration is longer than the thermalization time constants
of electrons and lattice and ablation proceeds as a thermal process.  If the laser intensity exceeds a threshold of the
order of $10^8$~W/cm$^2$ (corresponding to an energy density in the range of $0.1$~J/cm$^2$), the target surface is locally molten and the material partly
vaporized \cite{willmott,chichkov}. Above threshold, the ablation layer thickness quickly reaches the range of 10~nm per laser pulse. The resulting plume
contains atoms, molecules, and clusters as well as electrons and ions. Electrons are heated  by inverse bremsstrahlung in the interaction with
the laser light, leading to further ionization and heating of the ions. Finally, ions leave the plume away from the target surface at velocities in
the range of $10^4$~m/s (see e.g. Refs. \cite{thestrup,laska}). It has been shown that the distribution of ions over the charge states depends on the laser intensity and shifts to higher charges for higher intensity \cite{laska}.

In order to investigate the suitability of this process for ion production of relevant
elements with a low-power nitrogen laser, we use a time-of-flight (TOF) mass spectrometer to measure the ion yield and the distribution over the charge
states. These results are described in sections 2 and 3. In a second experiment, described in sections 4 and 5, a linear Paul trap was used to trap Th$^+$ ions
produced by laser ablation and the loading efficiency and dependence on the RF phase was investigated.


\section{Laser source and time-of-flight spectrometer}\label{sec:lasersource}

\begin{figure}[htb]
   \centering
   \includegraphics[width=0.95\columnwidth]{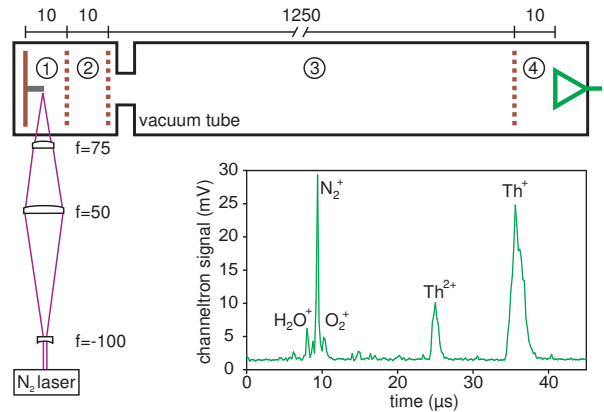}
   \caption{Setup of the time-of-flight mass spectrometer and typical experimental result. All distances are given in mm.
   1~-~ablation region, 2~-~acceleration region, 3~-~drift region, 4~-~detection region with channeltron. Dashed lines indicate grids. The inset shows a TOF spectrum obtained in the ablation of thorium metal.}
   \label{fig:setup}
\end{figure}

A commercially available nitrogen laser (model SRS NL-100) is used at a wavelength of 337 nm with a maximum pulse energy of 170~$\mu\textrm{J}$, a  pulse duration of 4~ns, and a maximum repetition
rate of 20 Hz. To obtain intensities higher than the ablation threshold, the transverse multimode laser beam was focused to a spot of dimensions ${100\,  \cdot 150\, \mu \textrm{m}}^2$ (to within $\pm10$\% uncertainty) using a lens system as indicated
in figure~\ref{fig:setup}. Thus the obtained peak intensity is  about $280\, {\textrm{MW}}/{\textrm{cm}^2}$ and the energy density is $1.1\, {\textrm{J}}/{\textrm{cm}^2}$.

In order to characterize the distribution of charged ablation products we use a TOF mass spectrometer \cite{wiley}. This technique is well suited here because the ions are produced during a very short time interval. If, subsequently, the ensemble of ions is accelerated in a static electric field to the same kinetic energy, different masses~$m$ and charges~$q$ will have a distribution of velocities which is proportional to $\sqrt{q/m}$.
The flight path of the ions is divided into four regions as outlined in figure~\ref{fig:setup}. At first, ions are produced in the ablation region. The ion cloud is then moved by a small electric field  to
the acceleration region where the ions  gain most of their kinetic energy. They are injected into the field-free drift region where the mass-to-charge resolution is obtained and are finally detected according to their arrival time.

The ablation targets were mounted in a cubic vacuum chamber segment that was evacuated to $10^{-6}\, \textrm{Pa}$. The sample is in electrical contact with a repeller plate held at a potential of +2000~V, defining
the ablation region with a spacing of 10~mm to the first grid at a voltage of +1900~V. A second, grounded grid, mounted 10~mm behind the first grid, creates the acceleration region. Attached to the cubic chamber, a 1.25~m long tube provides the drift region between the second grid
and a third grid, also at ground potential, at the end of the tube. The positive ions are then accelerated towards the cathode of a  channeltron electron multiplier that was held at a voltage of $-2000$~V. The signal current of the channeltron was
converted to a voltage using a transimpedance amplifier and was measured with a digital storage oscilloscope. A small fraction of the ablation laser
beam was detected by a photodiode in order to start the oscilloscope scan.

A typical result for the ablation of thorium is shown in figure~\ref{fig:setup}. It shows a signal of Th$^+$ with the maximum of the peak in good agreement with the calculated flight time. A Th$^{2+}$ signal is clearly seen as well, but no evidence of Th$^{3+}$ at
the expected time of $\approx19\,\mu$s is found. 
The width in the arrival times for these ions is much larger than the laser pulse duration and is a consequence of field inhomogeneity in our TOF setup. In addition, an unresolved contribution from ThO$^+$ ions contributes to a shoulder trailing the Th$^+$ signal. 
Much narrower signals of light molecular ions from the residual gas or desorbed from the surface are observed at $\approx 10\,\mu$s and show similar distributions in ablation measurements for
all tested materials. The peaks can be assigned to molecular ions of water, nitrogen, and oxygen.


\section{Ion yield and distribution of charge states}\label{sec:ionyield}

\begin{figure}[htb]
   \centering
   \includegraphics[width=0.95\columnwidth]{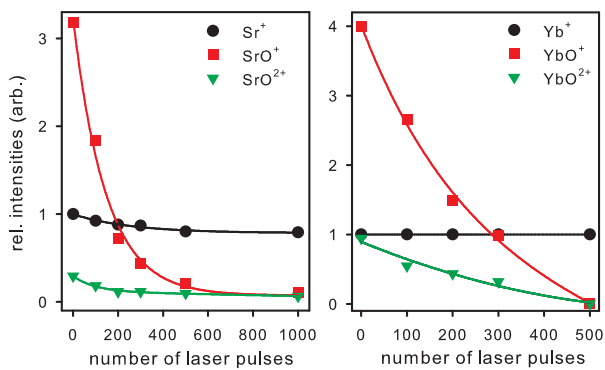}
   \caption{Evolution of the yield of molecular and atomic ions from laser ablation of oxidized samples of Sr (left) and Yb (right) as a function of the number of laser pulses applied to the same surface area. The lines superimposed on the data points are a guide to the eye.}
   \label{fig:strontium_and_ytterbium_cleaning}
\end{figure}

A number of elements that are of interest for current ion trap experiments were tested and the ablation yields for the observed singly and doubly positively charged ions are
noted in Table~\ref{tab:relative_ablation}. The number of ions was calculated from the pulse area of the channeltron signal and was normalized to Th$^+$, corresponding to a detected ion number of about $10^5$ per laser pulse. From all
tested metals, including the refractory metals molybdenum, tantalum, tungsten, and also from a silicon crystal, singly charged atomic ions could be produced effectively. Doubly charged ions were observed for Mo and Th only.  For samples of Ca, Sr and Yb,
oxide molecules could be detected as well. Since the yield of the oxide ions was not stable for the investigated metals as seen in
figure~\ref{fig:strontium_and_ytterbium_cleaning} and discussed below, the detected amount of oxides is not included in the table. In general, the total number of detected ions scales roughly inversely proportional with the heat of evaporation of the corresponding metal.

\begin{table}[htb]
   \centering
   \caption{Relative ablation ion yields for different elements and ionization states at a laser intensity of 250~MW/cm$^2$, normalized to the yield of Th$^+$.}
   \label{tab:relative_ablation}
      \begin{tabular}{lccccc}
      \hline\noalign{\smallskip}
         element & symbol & 1+ & 2+ \\
      \noalign{\smallskip}\hline\noalign{\smallskip}
         aluminum & Al & 0.5 & \\
         silicon & Si & 0.8 & \\
         calcium & Ca & 4.7 &\\
         copper & Cu & 1.1 & \\
         strontium & Sr & 1.4 &\\
         molybdenum & Mo & 0.6 & 0.1\\
         indium & In & 1.8 &\\
         ytterbium & Yb & 0.8 &\\
         tantalum & Ta & 0.7 &\\
         tungsten & W & 0.4 &\\
         gold & Au & 0.6 &\\
         thorium & Th & 1.0 & 0.3\\
      \noalign{\smallskip}\hline
      \end{tabular}
\end{table}

The TOF spectra detected from Sr and Yb samples showed a significant abundance of oxide ions. The employed metal samples were heavily oxidized due to exposure to air
for prolonged periods. Usually, these metals must be handled in a protective gas atmosphere to prevent the formation of oxide
layers.
Right after focussing the laser on the untreated surface of the Sr sample, the amount of detected SrO$^+$ ions is considerably  higher than the amount of Sr$^+$ ions. Doubly charged SrO$^{2+}$ ions
were detected as well.
Repeated laser pulses on the same spot decrease the amount of detected SrO$^+$ and SrO$^{2+}$ ions significantly until much less molecular ions than atomic Sr$^+$ ions are detected after 1000 shots. Figure~\ref{fig:strontium_and_ytterbium_cleaning} shows the number of oxide and atomic  ions in the produced ablation plasma as a function of the number of
laser pulses on the same spot for two samples of Sr and Yb. For both elements, laser ablation turns out to provide an effective method to remove the oxide layer.
The approach of cleaning the surface with multiple laser ablation pulses prior to an actual experiment is an effective method to load
ion traps with highly reactive elements without prior handling in a protective gas atmosphere.

\section{Loading of a linear Paul Trap}\label{sec:loading}

In order to test the suitability of the method for loading an ion trap, the laser and optics shown in Fig. 1 were used together with a linear RF Paul trap and the ablation beam was focussed on a target located between the trap electrodes. A segmented linear Paul trap was
constructed to store a high number of ions. It consists of cylindrical rods with a diameter of 8~mm, mounted at a diagonal distance of 7~mm between the
surfaces of the rods as it is shown in Fig. 3. The trap electrodes are made of copper-beryllium and consist of three parts, where the middle
section for storage of the ions is 20~mm with an overall length of the three segments of 160~mm. The experiments were conducted with Th$^+$ ions.

The operation principle of a Paul trap relies on an alternating electric quadrupole field. The equations of motion for an
ion with mass $m$ and charge $e$ in a segmented linear Paul trap can be described by a set of Mathieu equations~\cite{paul}. The
stability of the radial confinement is determined by the dimensionless parameters:
\begin{align}\label{eq:stab_param}
a_x=\frac{4e}{m\Omega^2}\frac{U_{DC}}{r_0^2}-\frac{4e}{m\Omega^2}\frac{\kappa U_{EC}}{z_0^2}\, , && q_x&=\frac{2e}{m\Omega^2}\frac{U_{RF}}{r_0^2}.
\end{align}
Here $U_{RF}$ is the amplitude of the RF voltage (at angular frequency $\Omega$), $r_0$ and $z_0$ are the radial and axial
distances from the trap center to the end of the middle electrode.
$U_{DC}$ is a static voltage that may be applied to two electrodes of the middle section in order to lift the degeneracy of the radial secular frequencies.  $U_{EC}$ is the static voltage that is applied to all electrodes of the two outer sections for the axial confinement within the middle section with the geometrical factor $\kappa\approx 0.3$. In the experiments reported here a frequency $\Omega \approx {2 \pi}\cdot 2\ \textrm{MHz}$ and $U_{RF}$ between 400~V and 900~V were used, resulting in a $q$-parameter in the range 0.2--0.4.

The ablation target is placed in the central position between the two lower trap electrodes (see Fig. 3).
The ablation laser beam passes through the gap between the upper electrodes and is focussed on the target. The grounded ablation target produces additional gradients of the trapping potential, with the effect of repelling the ions along the axis. For ions in the middle section at an axial distance of about 80~mm from the target, this perturbation has no detectable influence. The yield of
captured ions depends on the distance of the ablation target from the axis of the trap. It could be increased by about a factor of two by moving the target 1~mm higher, i.e. closer to the trap axis, but with the disadvantage of significantly higher shot-to-shot fluctuations of the ion number. With the target closer to the trap axis, the ions are produced in a region of smaller trap potential, but the trap is also perturbed more strongly by the target. To maintain a high loading efficiency the position of the focus of the ablation laser beam on the target is shifted periodically because the ablation induces changes in the surface profile.

\begin{figure}[htb]
   \centering
   \includegraphics[width=0.50\columnwidth]{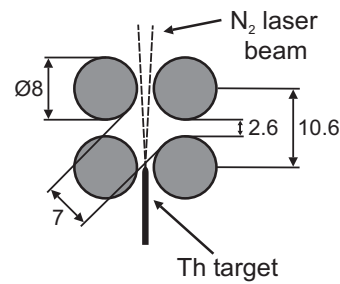}
   \caption{Schematic of trap electrodes and Th sample. The Th target is installed in the central position between the two lower trap electrodes of the loading section. The ablation laser beam passes through the gap between the upper electrodes. Units are in mm.}
   \label{fig:electrodes and sample}
\end{figure}

The trap is installed in a low permeability stainless steel vacuum chamber with a diameter of 300~mm. Two CaF$_2$ windows in
the top and the bottom of the chamber are facing the middle section for the observation of fluorescence light emitted by the ions.  A fused silica window in the top provides access for the ablation laser and additional windows around the circumference of the chamber pass laser beams for excitation of the trapped ions. To maintain a pressure in the range of $10^{-8}$ Pa, a turbomolecular pump is supported by a NEG (Non-Evaporable Getter) pump that  removes water, oxygen and hydrogen but does not pump noble gases that will be used for collisional cooling of the ions. A buffer gas system is attached to provide a supply of helium of high purity.

Different methods were applied to estimate the number of trapped ions. A nondestructive and quantitative method uses electronic detection of the mirror current induced in one of the trap electrodes when the ion cloud is resonantly excited to a collective oscillation at the secular frequency \cite{fischer}. A lower limit of $10^5$ stored ions loaded after 5--10 ablation pulses was deduced from this signal.

A simple quantitative method to measure the ion number is the counting of ions with a channeltron after release from the trap. After a loading and trapping period, the amplitude of the RF voltage is reduced until the ions are extracted towards the channeltron. The opening of the channeltron was facing towards the middle section of the trap perpendicular to the trap axis from a distance of 5~cm. These experiments were performed at buffer gas pressures in the range of $10^{-4}$~Pa, limited for the operation of the channeltron without discharges.   Typically, on the order of $10^4$ ions were detected after one ablation pulse, consistent with the result from electronic detection described above.  The absolute efficiency of this detection method is uncertain because ions leave the trap not only in the direction towards the channeltron. The main advantages of the method turned out to be the experimental simplicity and its insensitivity with respect to the ion species to be detected. This method was used for the experiments on the RF phase dependence of the loading efficiency that are described in section 5.

\begin{figure}[htb]
   \centering
   \includegraphics[width=0.95\columnwidth]{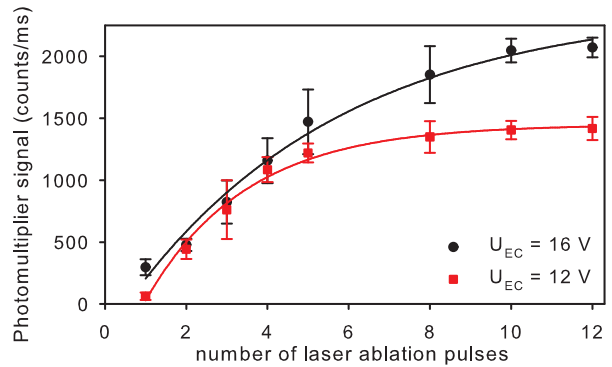}
   \caption{Fluorescence signal of stored Th$^+$ ions for multiple ablation pulses with different DC axial confinement voltages, showing cumulative loading of the trap.}
   \label{fig:loading_characteristics}
\end{figure}

Future experiments with this setup will focus on laser excitation of the stored Th$^+$ ions and optical fluorescence detection provides a convenient measure of the relative ion density in the center of the trap.  The ions were excited with a tunable extended-cavity diode laser on the strongest resonance line at 401.9~nm which connects the (6\emph{d}$^{2}$7\emph{s})\emph{J}=3/2 ground state with the (6\emph{d}7\emph{s}7\emph{p})$ \linebreak $\emph{J}=5/2 state at 24874~cm$^{-1}$.  Helium buffer gas at a pressure of $0.2$~Pa was used for collisional cooling and quenching of metastable states.  The fluorescence signal was detected at the same wavelength as the excitation with a photomultiplier and was recorded as a function of the number of ablation laser pulses. Figure~\ref{fig:loading_characteristics} shows the results for two different DC axial confinement voltages $U_{EC}$. The experiment shows that laser ablation loading in this trap geometry is cumulative over several ablation shots before the ion number saturates. The saturation ion number depended on the trap parameters and but also on the ablation conditions like the laser pulse energy.

\section{Investigation of the RF phase dependence of the loading efficiency}\label{sec:phase}

Theoretically, the trajectory of an ion in a RF trap is described by a solution of the Mathieu equation. The stability of the
solution (i.e. whether the distance of the ion from the trap center stays finite for all times) depends only on the values of the parameters $a$ and
$q$ (see Eq. (1)). For a real ion trap with a given extension of the trapping volume that is limited by the electrode surfaces, the stability of a trajectory
depends on the initial position and velocity of the ion and on the initial phase of the RF field as well. This phase dependence of ion loading has  already been studied theoretically in early works, for the rotationally symmetric Paul trap with ions created at rest inside the trap~\cite{fischer}, and for ions injected through holes in the electrodes or through the gap between electrodes \cite{schuessler1,schuessler2}. In both cases the
phase-dependence is  found to be pronounced especially for high values of $q$. For an initially homogeneously filled trap volume, Ref. \cite{fischer} predicts a modulation of the capture efficiency with the
RF phase between 100\% and 50\% for $q=0.28$.  Very few experimental studies of this effect have been reported
\cite{hemberger,robb,vargas}, because the most commonly used method of ion creation via electron impact requires a time that is much longer than the
period of the RF field.

\begin{figure}[htb]
   \centering
   \includegraphics[width=0.95\columnwidth]{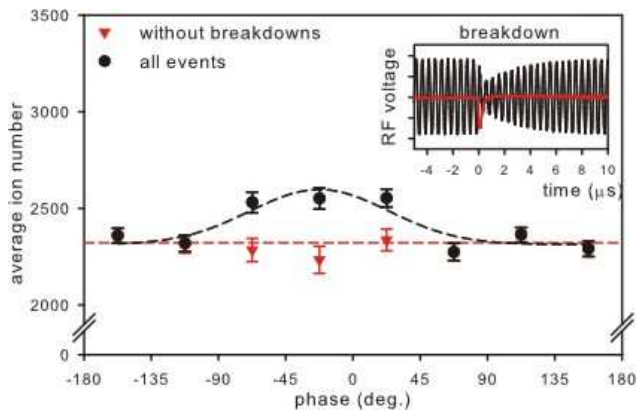}
   \caption{Trapped ion number as a function of the timing of the ablation pulse with respect to the RF phase. The triangles show data for events with constant RF amplitude, the black circles include events when the RF amplitude was momentarily reduced (see inset) because of a discharge breakdown that was ignited by photoelectrons from the ablation pulse. The connecting curves are drawn as a guide to the eye.}
   \label{fig:breakdown}
\end{figure}

In the experiment reported here, laser ablation loading of a linear Paul trap produces ions on the nanosecond timescale and is therefore capable of
resolving a phase dependent loading efficiency. Since the pulse timing of the ablation laser showed a considerable jitter of about $3\,\mu$s, it was
not possible to trigger it on a definite value of the RF phase. Instead, the phase of the RF field was measured for every ablation shot and the
observed ion number on the channeltron was stored and averaged over 45$^\circ$ wide bins. 
The phase convention $U_{RF}\propto \cos \Omega t$ was used here, i.e. maxima of the RF field strength appear at phases $0$ and $\pm 180$ degree.   
The $q$ value was chosen in the range that provides the most efficient trapping conditions.
The result is shown for trap parameters
$q=0.4$ and $a=0$ as the lower-lying curve (triangles) in figure~\ref{fig:breakdown}. Measurements for $q=0.2$ and $q=0.3$ lead to similar results. With a statistical uncertainty that allowed to resolve a relative change of the ion number of 4\%, no significant phase dependence of the loading rate was observed for these parameters.

In some events close to the field maximum at phase $0$, the ablation pulses were accompanied by a temporary  breakdown of the
RF field. In these events, the loading efficiency was about 20\% higher than in the case without breakdowns. The black circles in figure~\ref{fig:breakdown} show data that includes those events. This observation can be understood as follows: Free electrons produced by the
ablation laser pulse lead to the ignition of a discharge between the trap electrodes and the grounded ablation target. In these events the RF voltage
decreased to less than 50\% of its steady state value and recovered within about 5--10~cycles of the driving frequency or about 3--5~$\mu$s. A typical
time dependence of $U_{RF}$ during a breakdown is shown in the inset of figure~\ref{fig:breakdown} together with the signal from a photodiode that
detects the ablation pulse.  Similar effects have been described in \cite{kwong1,hashimoto}. The ignition of the discharge did not appear in a
completely deterministic way and evidently showed a dependence on the RF phase. It is initiated with highest probability when the trapping voltage is close to
its maximum and in one polarity only, i.e. once during the RF cycle. In order to test whether the increased loading efficiency during the RF
breakdown was associated with the time dependence of the RF amplitude (as proposed in \cite{hemberger}), a fast voltage controlled attenuator was used to
produce a similar envelope of the RF amplitude, correlated with the laser ablation pulse but without the ignition of a discharge. In this
test, no enhancement of the ion loading rate was observed. We conclude that the increase of the trapped ion number that was associated with the
discharge was due to further electron impact ionization of neutral Th within the ablation plume.

\begin{figure}[htb]
   \centering
   \includegraphics[width=0.95\columnwidth]{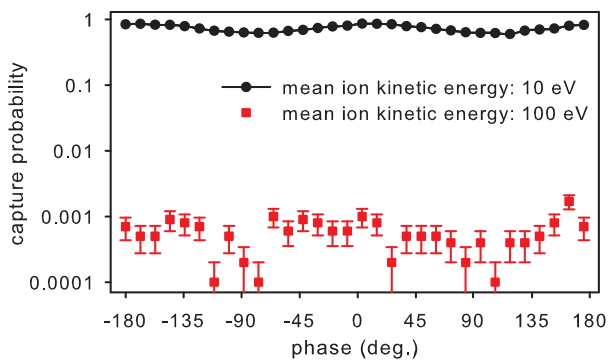}
   \caption{Results of a numerical simulation of the trap loading as a function of RF phase for the experimental parameters of Fig. 5 (lower data points). The upper curve is obtained for an ion distribution with ten times reduced initial kinetic energy.}
   \label{fig:numeric_distribution}
\end{figure}

In order to obtain a more complete understanding of these observations, a numerical simulation of the loading process was undertaken. The trap was modeled as infinitely long and ion trajectories were calculated in two dimensions in the plane orthogonal to the trap axis. The initial
velocity distribution of the ions in the expanding ablation plasma was derived from an experiment \cite{thestrup} performed at laser
parameters that were similar to the ones used here, indicating a mean kinetic energy of 100~eV. This is in agreement with observations based on the TOF experiments described in Sections 2 and 3. Simulations were performed for this velocity distribution and for one with the kinetic energy reduced by a factor of 10, to a value below the trap depth. The initial position distribution was assumed to be a point, corresponding to the tip of the ablation target. The simulation did not include any damping of the motion.
Collisional damping from helium buffer gas in the investigated pressure range of $10^{-4}$~Pa results in a damping time in the range of seconds \cite{schwarz}, i.e. in the range of
$10^6$ RF cycles.

The lower-lying data points in Figure 6 show the calculated capture probability for parameters identical to Fig. 5, obtained from the integration of individual ion trajectories with randomly chosen initial velocity for each value of the phase. The low average capture probability obtained here is in disagreement with the experimental finding of comparable ion numbers in the TOF and trap experiments. 
For the slower ion distribution (upper curve in Fig. 6), the maximum capture probability approaches unity and a phase-dependent modulation of about 20\% is obtained in the simulation. A similar modulation can be seen for the more realistic velocity distribution, though at a lower statistical significance, because of the small overall capture probability.

In comparison with the results from the experiments, it can be seen that the simulation underestimates the capture probability and predicts a phase dependence that is not observed experimentally.
Since the initial density of neutral and charged particles in the ablation plume is high \cite{willmott}, velocity changing collisions and
Debye shielding may be expected to play an important role in the loading of the trap.
 Shielding will reduce the influence of the trapping field on some ions as long as the density is high. A collision between atoms of identical mass can lead to strong momentum transfer. Both effects may therefore produce a sub-ensemble of ions close to trap center at a kinetic energy that is lower than expected from the model with noninteracting particles. Inclusion of these effects would make a simulation of the process much more involved. A RF phase dependence of the loading efficiency that also
depended on the ion density has been observed in an experiment with organic molecules \cite{robb}.

\section{Conclusion}

In conclusion, we have presented a study of the production of metal ions by laser ablation with a compact, low-power nitrogen laser. Significant numbers of singly charged ions could be produced for all the elements studied here. These were selected because they are of interest for ion trap experiments in the domains of optical clocks and quantum information processing. The method also proved to be effective for refractory metals and for thorium. Since ion production occurs on a nanosecond time scale it was possible to investigate the theoretically predicted dependence of the loading efficiency on the phase of the RF field. Contrary to this prediction that was derived for individual ions,  no significant phase dependence was observed in the experiment, presumably because of collective processes within the dense ablation plume.

Based on a rough estimate of the vaporized volume of $10^{-11}$~cm$^3$ of metal, i.e. about $10^{11}$ atoms, and a detected ion number below $10^5$ per ablation pulse, an overall ion production and trapping efficiency of order of magnitude $10^{-7}$ to $10^{-6}$ can be deduced for the experiments reported here. 
This may be compared to an efficiency of about $10^{-8}$ to $10^{-7}$ that has been reported for the loading of Th$^+$ ions from a heated wire source \cite{kaelber}. 

As mentioned in the introduction, the production of free electrons can be a critical factor for the operation of the ion trap because of stray charges deposited on insulators. With a work function of 3.5~eV for Th it can be estimated that both, photoelectron emission and thermal electron emission contribute and that an order of magnitude of $10^{10}$ electrons will be produced per pulse. This number is significantly lower than those typically used in loading from an atomic beam with an electron gun and it could be reduced further by using an ablation laser with a photon energy below the work function.

The efficiency in our laser ablation experiments seems to be primarily limited by the low fraction of charged ablation products and by the high kinetic energy of the ion cloud that expands from the ablation plume. A further optimization of the loading efficiency should therefore include a study of the laser heating of the plasma. A significant gain may also be expected by combining laser ablation with photoionization close to the center of the trap that will produce additional ions from ablated neutral atoms.


\begin{acknowledgements}
We thank Chr. Tamm for helpful discussions and D. Griebsch and Th. Leder for their expert technical support. This work was partially supported by DFG
within the cluster of excellence QUEST. OAHS acknowledges support from ITCR, MICIT and DAAD.
\end{acknowledgements}


\end{document}